\documentclass[10pt,a4paper]{article}
\usepackage{graphics}
\usepackage{amsmath}
\usepackage{amssymb}
\usepackage{amscd}
\usepackage[normal,small]{caption}
\usepackage{afterpage}
\usepackage{float,times}
\usepackage{subfigure}
\usepackage{rotating}
\usepackage{multirow}
\usepackage{fancyheadings}
\usepackage{epsfig}
\usepackage{cite}
\usepackage{theorem}
\usepackage{moreverb}
\usepackage{euscript}
\usepackage{psfrag}
\title{Generating the Baryon Asymmetry of the Universe in Split Fermion Models}
\author{Andrew Coulthurst\thanks{acoul@physics.unimelb.edu.au}}
\date{\small School of Physics, Research Centre for High Energy
Physics,\\The University of Melbourne, Victoria 3010, Australia}

\def\GeV{\textrm{ GeV}}
\def\spinor#1#2{\left(\begin{array}{c}#1\\#2\end{array}\right)}
\begin{document}
\maketitle
\begin{abstract}
The origin of the matter-antimatter asymmetry of the universe is one of the major unsolved problems in cosmology and particle physics.  In this paper, we investigate the recently proposed possibility that split fermion models -- extra dimensional models where the standard model fermions are localized to different points around the extra dimension -- could provide a means to generate this asymmetry during the phase transition of the localizing scalars.  After setting up the scenario that we consider, we use a single fermion toy model to estimate the reflection coefficients for scattering off the phase boundary using a more realistic scalar profile than previous work resulting in a different Kaluza Klein spectrum.  The value we calculate for $n_B/s$ is consistent with the mechanism being the source of the baryon asymmetry of our universe provided the $B-L$ violating processes have an efficiency of order $10^{-5}$.
\end{abstract}

\section{Introduction}

It is clear, at least in our neighborhood of the universe, that matter is far more common than antimatter.  Observations seem to rule out the possibility that the universe is made of separate regions of matter and antimatter domination due to the non-detection of annihilation photons from the boundaries between the regions \cite{astroph9707087}. Further, there is no known mechanism which would lead to such an ordering.  It thus appears that there is far more matter than antimatter in the universe, or equivalently that the universe has a net baryon number, $B$.  The origin of this baryon asymmetry of the universe (BAU) is one of the major unanswered questions facing cosmology and particle physics.  The standard way to parameterize the size of the asymmetry is as the ratio of the universe's  baryon number density to its entropy density:
\begin{equation}
n_B/s\sim 10^{-10}.
\label{eqn:BAU}
\end{equation}

Given this, we require a theory which can generate a baryon asymmetry from an initially baryon symmetric universe.  Sakharov \cite{Sakharov} showed there were three conditions necessary for a theory to create such an asymmetry: 1) Baryon number violation, 2) C and CP violation, and 3) Departure from thermal equilibrium\footnote{It should be noted that none of Sakharov's conditions are strictly necessary for baryogenesis to occur and models have been developed where each can be circumvented \cite{Dolgov:1991fr}.  However, such models have thus far had limited success and at least for the type of models we consider here, all three conditions are necessary.}. It is well known that these three requirements may be met by the Standard Model (SM) during the Electroweak Phase Transition (EWPT) via a mechanism known as electroweak (EW) baryogenesis \cite{hepph9803479} but unfortunately the parameters of the SM -- in particular the values of parameters in the CKM matrix and the mass of the Higgs Particle -- conspire to make the resulting asymmetry many orders of magnitude too small to explain that which we observe.  However, it should be noted that the basic mechanism of EW baryogenesis -- the CP non-invariant scattering of fermions off the boundary between two phases during a phase transition -- remains a feasible means to generate the asymmetry if we can find an alternative context in which it might take place.  As we will show it is possible that models with split fermions may provide such an alternative context.

Split fermion models are extra dimensional models in which the SM fermions are confined to a brane by one or more localizing scalar fields which have a non-trivial vacuum expectation value (VEV) along the extra dimension.  (Throughout we will assume, for simplicity, that there is only a single extra dimension.)  However, unlike the standard scenario the position along the extra dimension to which the fields are localized varies for the different quark and lepton flavors leading to a so-called fat brane structure.  Such models are attractive since they can provide a natural explanation for both the hierarchy in the SM Yukawa couplings and the suppression of proton decay with both these features of the SM resulting from the exponentially small overlap of wavefunctions of fermions localized to different points.

As was noted in \cite{hepph0401070,hepph0505222}, such a scenario also provides an arena where we might hope to meet all Sakharov's conditions and where baryogenesis may have occurred.  This would have taken place during the phase transition where the localizing scalar fields acquire VEVs.  In analogy to the EW baryogenesis model, during this period there are regions of both broken and unbroken phase.  In the unbroken phase, where the scalars have no VEV, the SM fermions are unlocalized and the theory is fully 5d.  In this phase the fermion wavefunction overlaps are large and we expect Yukawa couplings to naturally be $\mathcal{O}(1)$ and baryon number violating processes to proceed rapidly.  In the broken phase, the SM fermions are localized and the effective theory is the 4d SM -- Yukawa couplings are hierarchical and baryon number violating processes (with the exception of sphaleron processes) are exponentially suppressed.  Thus both B and CP violation can be large in the broken phase and we may have a departure from thermal equilibrium provided the phase transition is first order. 

To test if this model is feasible involves calculating the CP non-invariant scattering of SM fermions off the phase boundary and determining the size of the asymmetry in the reflection of particles and anti-particles.  However, as is generally the case when we consider fields propagating in compact extra dimensions, the problem is complicated by the presence of Kaluza-Klein (KK) modes of the SM particles.  Further, the additional scalars we must add to the theory in order to localize the fermions introduce numerous, relatively unconstrained parameters.  In this paper we therefore assume that the asymmetry generated in a realistic three generation model may be approximated by a calculation in a simplified toy model with just a single fermion.  This approach has previously been considered to estimate the BAU generated in split fermion models \cite{hepph0401070,hepph0505222} with results consistent with this mechanism being the source of the BAU. However, these papers consider a simplified profile for the localizing scalar leading to a markedly different KK spectrum in the broken phase than that found in a detailed study of this spectrum with a more realistic scalar profile \cite{hepph0309115}.  Since the reflection asymmetries depend crucially on the KK spectrum in the two phases (including off-shell modes), this has the potential to dramatically alter these previous results.\footnote{\cite{hepph0003312} and \cite{hepph0112360} also use split fermion models to generate the BAU but via very different mechanisms to that we consider here.}

The outline of this paper is as follows:  In Section \ref{sec:EWB} we give a brief review of electroweak baryogenesis while in Section \ref{sec:AS} we introduce the original split fermion setup.  We outline the details of our implementation of split fermion baryogenesis in Section \ref{sec:cosmo} and describe the simplified single fermion model and the KK modes it will contain in Section \ref{sec:toymodel} (with some necessary results from a model with an infinite extra dimension contained in Appendix \ref{sec:infsolns}).  In Section \ref{sec:coeffs} we calculate the reflection coefficients for this model and find the resulting asymmetry before concluding in Section \ref{sec:conclusion}.

\section{Electroweak Baryogenesis}
\label{sec:EWB}

In the following paragraphs, we give a very brief outline of the electroweak baryogenesis process, largely as an analogy to the split fermion baryogenesis mechanism which we will consider for the remainder of the paper.  For a more thorough review see \cite{hepph9803479} and references contained therein.

EW baryogenesis attempts to generate the BAU using only SM physics.  All three of Sakharov's conditions may be met: 1) Baryon number is violated by non-perturbative sphaleron processes \cite{'tHooft:1976up}, 2) CP violation occurs due to a non-zero phase in the CKM matrix, and 3) there may be a departure from thermal equilibrium provided the electroweak phase transition is first order.  Above the temperature of the EWPT, $T_{\textrm{EWPT}}$, the Higgs particle has zero VEV and sphaleron processes may proceed rapidly.  Once the universe cools to $T_{\textrm{EWPT}}$, regions of broken phase, in which the Higgs has non-zero VEV and sphaleron processes are exponentially suppressed, begin to appear.  As the universe cools further these bubbles of broken phase expand and envelop surrounding regions until they fill the visible universe.  As the bubbles expand, quarks from the unbroken phase scatter off their walls.  Due to CP non-invariance, this scattering is different for quarks and antiquarks and their reflection coefficients, $R_{\textrm{RL}}^{fi}$, $\bar R_{\textrm{RL}}^{fi}$ (representing the scattering of a left-handed quark of flavor $i$ into a right-handed quark of flavor $f$, and its CP-conjugate process, respectively), will differ.  As a result, a net baryon number will build up inside the bubble with an equal and opposite number outside.  Recall, however, that B violation proceeds rapidly outside the bubble so this asymmetry is quickly wiped away while inside the bubble B violation is suppressed, allowing the asymmetry here to be preserved.  Thus, once the bubbles expand to encompass the visible universe, we may be left with the required matter-antimatter asymmetry.  Ignoring the effect of particles from the broken phase scattering back into the unbroken phase, the resultant asymmetry is given by:
\begin{equation}
n_{\textrm{B}}=\int\frac{dE}{2\pi}(n^u_L(E)-n^u_R(E))\Delta(E),
\label{eqn:EWn}
\end{equation}
where $\Delta(E)=\textrm{Tr}(R_{\textrm{RL}}^\dag R_{\textrm{RL}}-R_{\textrm{LR}}^\dag R_{\textrm{LR}})$, and $n^u_{\textrm{R,L}}$ are the Fermi-Dirac distributions in the unbroken phase boosted to the wall frame.  Taking the wall speed to be $v_w\approx 0.1$, we may expand this expression to first order in $v_w$ obtaining \cite{hepph9404302}:
\begin{equation}
\frac{n_{\textrm{B}}}{s}\approx\frac{10^{-3}}{T_{\textrm{EWPT}}}\int\frac{dE}{2\pi}n_0(E)(1-n_0(E))\frac{(\vec p_{\textrm{L}} -\vec p_{\textrm{R}})\cdot \hat v_w}{T_{\textrm{EWPT}}}\Delta(E),
\label{eqn:EWnpert}
\end{equation}
where $n_0(E)$ is the rest frame Fermi-Dirac distribution function and $\vec p_{R,L}$ are the particle momenta.

Hence to calculate $n_B/s$, we need to find values for the reflection coefficients.  This is done by writing down wavefunctions in each phase for particles scattered off the bubble wall in terms of reflection and transmission amplitudes.  If we assume the wall is thin (i.e. $\delta_w<v_w\tau_i$ where $\delta_w$ is the wall thickness and $\tau_i$ are the timescales of processes taking place), continuity requires the wavefunction in the two phases to match on the wall.  Enforcing this requirement allows us to find the reflection/transmission amplitudes and in turn the reflection coefficients and the resultant BAU.

While this model of baryogenesis is attractive in that it does not require any new physics (beyond the as yet undetected Higgs particle), it is now ruled out, at least in its simplest formulation.  CP violation in the SM is a small effect as parameterized by the Jarlskog parameter, $J\sim 10^{-5}$ \cite{PDBook} and there has been disagreement as to whether it is sufficient to generate the required asymmetry once quantum effects are fully included \cite{hepph9305274,hepph9312215,hepph9404302}.  More categorically, the Higgs mass is now restricted to be greater than $114 \GeV$ \cite{PDBook}.  Such a large Higgs mass leads to a second order EWPT \cite{Shaposhnikov:1986jp,Shaposhnikov:1987tw} preventing the required departure from thermal equilibrium.  Further, with a large Higgs mass, the suppression of sphaleron processes below $T_{\textrm{EWPT}}$ is reduced and these processes can proceed at a rate sufficient to wipe out any asymmetry that is generated.  We emphasize that these are problems related to the parameters of the SM and are not intrinsic to the mechanism described.  It is therefore conceivable that the asymmetry is generated during a phase transition where some scalar other than the SM Higgs acquires a VEV.  It is this possibility we explore in the context of split fermions.

\section{Arkani-Hamed Schmaltz Model}
\label{sec:AS}
\begin{figure}
\centering
\includegraphics[width=0.6\textwidth]{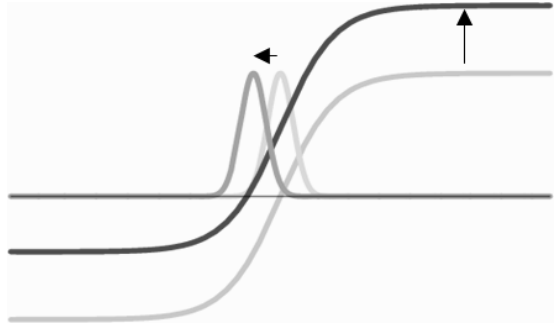}
  \caption{When it acquires a VEV, boundary conditions force the scalar to take a domain wall or kink configuration along the extra dimension.  In this environment the fermion solution is a `Gaussian' localized around the zero of the scalar. By giving a fermion a bare mass $m$, we can shift its wavefunction to be centred about $x_5=-m/2\mu$.  Giving the different fermions different bare masses allows us to split up the fermions along the extra dimension.}
  \label{fig:braneshift}
\end{figure}

In their original proposal \cite{hepph9903417}, Arkani-Hamed and Schmaltz (AS) considered a fermion and a scalar in an uncompactified 4+1d space with a Lagrangian:
\begin{equation}
\mathcal{L}=\bar{\psi}(x,x_5)(i\gamma^\mu\partial_\mu+i\gamma^5\partial_5+\phi(x,x_5)+m)\psi(x,x_5),
\label{eqn:ASlagrangian}
\end{equation}
where $x$ labels our normal 3+1d spacetime and $x_5$, the extra dimension. The profile of the scalar field's expectation value along the extra dimension is that of a domain wall centred at $x_5=0$.  Initially $m$ is set to zero.  For the case where the scalar profile is approximated by $\langle\phi(x_5)\rangle=2\mu^2x_5$ close to $x_5=0$, the Dirac equation for the fermion is found to have a chiral zero mode solution:
\begin{equation}
\psi_{L,0}(x_5)=\frac{\mu^{1/2}}{(\pi/2)^{1/4}}\exp(-\mu^2x_5^2),
\label{eqn:ASpsiL}
\end{equation}
while the corresponding right-handed solution is not normalisable.  This solution is a Gaussian centred at $x_5=0$. If the fermion is now given a bare mass i.e. $m\neq0$, we find the same left-handed chiral solution now centred about $x_5=-m/2\mu$ (Figure \ref{fig:braneshift}).  Thus if we generalize to the many fermion case we see we can obtain chiral fermions localized at different points along the extra dimension if we give these fermions different bare masses.

AS note two attractive properties of such a setup.  Firstly, we may consider SM fermions where the bare masses of the SU(2) singlets and doublets differ, leading to their separation by a distance $r$ along the extra dimension.  Adding a Higgs to the Lagrangian, if the fundamental 5d Yukawa coupling between the fermions and the Higgs is considered to be $\mathcal{O}(1)$, the effective 4d Yukawa couplings are found to be suppressed by the small overlap between the wavefunctions for the SU(2) singlet and doublet of a given particle. As a function of $r$ this suppression is found to be $e^{-\mu^2r^2/2}$.  Thus by considering fermions with a small range of bare masses we can naturally obtain the observed hierarchy in SM Yukawa couplings.

A somewhat more involved analysis shows that a similar separation between quark and lepton fields can lead to the almost complete suppression of dangerous operators in the 5d Lagrangian (such as $qqql$ terms) which would otherwise lead to proton decay occurring at unacceptable rates.

Since the work of AS, there have been numerous variations on their original setup and investigations into these models' phenomenology, experimental signatures and consistency with the SM \cite{hepph0003312,hepph0112360,hepph0007350,hepph0110126,hepph9909411,hepph9912265,hepph0012289,hepph0104312,hepph0210053,hepph0306193,hepph0309218,hepph0401070,hepph0505222,hepph0407260,hepph0401108}.  Most importantly, models where the extra dimension is compactified as a circle of radius $R$ have been considered \cite{hepph0007350,hepph0110126,hepph0210053}.  Such models require orbifold boundary conditions (OBCs) in order to reproduce chiral fermions and an alternative means to control the overlap between the fermions since bare mass terms are forbidden by the OBCs.  Such means include the introduction of non-renormalizable \cite{hepph0007350} or boundary \cite{hepph0110126} terms, non-uniform scalar-fermion couplings\cite{hepph0110126} or the addition of a second localizing scalar \cite{hepph0210053}.  We will consider the last of these options since it also has the potential to provide the source of CP violation which we require (see Section \ref{sec:cosmo}).

In the context of baryogenesis, this model is particularly suggestive.  It contains a scalar field with a non-zero VEV meaning there may have been a first order phase transition when this was acquired.  This could potentially provide us with the necessary departure from thermal equilibrium.  Prior to the transition, the SM fermions are unlocalized and there are no suppressions caused by small wavefunction overlaps.  Thus B violation may proceed rapidly and, due to the large Yukawa couplings, CP violation may be large.  It is therefore possible that all of Sakharov's conditions are met and we may be able to generate the BAU providing certain cosmological conditions are met as we discuss in the following section.
\section{Split Fermion Baryogenesis}
\label{sec:cosmo}

In order to satisfy Sakharov's conditions and generate the BAU, our split fermion model must contain CP violation. It has recently been proposed that it may not be necessary to include any new sources of CP violation and that the CKM phase could be all that is required to generate the BAU using split fermions \cite{hepph0505222}.  We do not consider this possibility here and instead follow \cite{hepph0401070} and introduce a second localizing real scalar to our Lagrangian:
\begin{equation}
\mathcal{L}\sim \bar \psi_i(f^{ij}_1\phi_1+f^{ij}_2\phi_2)\psi_j+V(\phi_1,\phi_2).
\label{eqn:Ltwoscalar}
\end{equation}
In a model with three fermion generations, the $f^{ij}_{1,2}$ will contain one physical complex phase, $\varphi$, which will lead to CP violation.  Since we anticipate all the $|f^{ij}_{1,2}|$ to be of order one, we expect the CP violation to be large.   As mentioned above, the addition of a second localizing scalar would also allow for the fermions to be localized to different points in the bulk, and it has been shown \cite{hepph0210053} that it is possible to reproduce the SM Yukawa couplings, although the choice taken there is not the only possibility and the $f$s are not uniquely determined.

Having set out the particle content of the model, the most important factor in determining its cosmological context is the temperature, $T_{\textrm{BAU}}$,  at which the localizing scalars acquire VEVs (for simplicity we assume this happens simultaneously for both $\phi_{1,2}$) leading to a phase transition and the possible generation of the baryon asymmetry.  General arguments suggest that to be consistent it is likely we require that $T_{\textrm{BAU}}\sim 1/\pi R$ \cite{hepph0401070}. Depending on the particular split fermion model considered and the regions of the parameter space explored, studies searching for lower bounds on $1/R$ (resulting from the need to suppress additional flavor changing neutral currents) find results ranging from a few TeV to a few thousand TeV \cite{hepph9911252,hepph0210414,hepph0306193}.  For our purposes, it is important that this suggests $T_{\textrm{BAU}}\gg T_{\textrm{EWPT}}$ and the Higgs has zero VEV during the epoch when we are considering baryogenesis to have taken place.   This means we are anticipating that the BAU will be generated well above the EWPT during an epoch when sphaleron processes which violate $B+L$ were proceeding rapidly.  As a result, if only $B+L$ were violated in our construction, any BAU generated at $T_{\textrm{BAU}}$ would have been completely wiped out by the time sphaleron processes were suppressed at the EWPT.  It is therefore necessary that the asymmetry is produced by $B-L$ violation.  The resulting asymmetry will clearly depend on the efficiency of this process, $\epsilon_{B-L}$.

The leading order baryon violating processes allowed by the SM gauge symmetries, from terms of the form $\frac{1}{M_*^3}QQQL$, where $M_*$ is the cut-off, conserve $B-L$ and are therefore not relevant in this context.  The lowest order processes which can violate $B-L$ have the symbolic form $\frac{1}{M_*^4}(QD_MQ)(Q\Gamma^M\overline L)$ where $D_M$ is the covariant derivative, $\Gamma^M$ are the gamma matrices in 5D and $M=0,1,2,3,5$.  In \cite{hepph0505222} it was shown that assuming the theory is strongly coupled at $M_*$, naive dimensional analysis (NDA) leads us to estimate that these processes proceed at a rate:
\begin{equation}
\Gamma_{B-L}\sim\frac{(4\pi)^4}{8\pi}\left(\frac{T}{M_*}\right)^8T.
\end{equation}
In order that the asymmetry generated inside the bubble by the CP non-invariant scattering off the wall is preserved, the opposite asymmetry outside the bubble wall must be washed-out by $B-L$ violating processes before the bubble expands to encompass that region of space.  Using $T_{\textrm{BAU}}\sim1/\pi R$, the efficiency of this process is therefore given by:
\begin{equation}
\epsilon_{B-L}\sim 2\Gamma_{B-L}t_w\sim\frac{1}{v_w}\left(\frac{1}{RM_*}\right)^8,
\label{eqn:epsilonB-L}
\end{equation}
where $t_w\sim 4/(Tv_w)$ is the timescale associated with the thin bubble wall \cite{hepph0505222,hepph9302210} and there is a factor of 2 since each interaction violates $B-L$ by 2 units.  We are unable to constrain this inefficiency any further since the split fermion mechanism effectively blinds us at low energies to the physics at energies above $T_{\textrm{BAU}}$ and in particular the value taken by $M_*$.  It is possible that this can be better constrained by cosmological considerations but such an analysis is well beyond the scope of this paper.

The setup is now analogous to that of EW baryogenesis described above.  Again the BAU will be generated by asymmetries in the scattering of quarks and antiquarks at the boundary between the regions where the scalars have zero and non-zero VEVs respectively.  By analogy with Equation \eqref{eqn:EWn}, we can write:
\begin{equation}
\frac{1}{\epsilon_{B-L}}\frac{n_{\textrm{B}}}{s}\approx\frac{10^{-3}}{T_\textrm{BAU}}\sum_{\substack{f,i \\ m,n}}\int\frac{dE}{2\pi}n_0(E)(1-n_0(E))\frac{(\vec p_{\textrm{L}}^{f^{(n)}} -\vec p_{\textrm{R}}^{i^{(m)}})\cdot \hat v_w}{T_\textrm{BAU}}\Delta^{f^{(n)} i^{(m)}}(E),
\label{eqn:npertmulti}
\end{equation}
Note that since we are working in an extra dimension we must sum over not only the flavors of the incoming and outgoing modes, but also their KK numbers, $n,m$.  This means
\begin{equation}
\Delta^{f^{(n)} i^{(m)}}=\left(R^{{f^{(n)} i^{(m)}}^\dagger}_{RL}R^{{f^{(n)} i^{(m)}}}_{RL}-R^{{f^{(n)} i^{(m)}}^\dagger}_{LR}R^{{f^{(n)} i^{(m)}}}_{LR}\right)\sim |R_{RL}^{{f^{(n)} i^{(m)}}}|^2\sin \varphi,
\end{equation}
and the asymmetry is difficult to calculate, even numerically. Aside from calculational difficulties, there is yet to be a dynamical realization of the split fermion idea which reproduces the Yukawa couplings of the SM while suppressing proton decay.  As a result, even if we could perform the calculation, we do not have a fully realistic model to peform it on.  We therefore simplify the problem by calculating the reflection coefficients in a model containing only a single fermion flavor. In doing this, we anticipate that they will carry over to the three generation case with corrections of $\mathcal{O}(1)$ \cite{hepph0505222}.  Having made this simplification, the equation for the asymmetry reduces to:
\begin{equation}
\frac{1}{\epsilon_{B-L}}\frac{n_{\textrm{B}}}{s}\approx\frac{10^{-3}}{T_\textrm{BAU}}\sum_{m,n}\int\frac{dE}{2\pi}n_0(E)(1-n_0(E))\frac{(\vec p_{\textrm{L}}^n -\vec p_{\textrm{R}}^m)\cdot \hat v_w}{T_\textrm{BAU}}\Delta^{nm}(E),
\label{eqn:npert}
\end{equation}
with $\Delta^{nm}(E)\sim |\mathcal{R}^{nm}|^2$ (where $\mathcal{R}^{nm}$ are the single fermion reflection coefficients) since we anticipate an order one phase.  Our problem is now reduced to finding the reflection coefficients.  To do this we must find the fermion wavefunctions for all KK modes, in both phases.  In this single fermion framework, neither of the qualitative effects of adding a second localizing scalar -- a non-removable complex phase and fermion splitting -- is present.  We are therefore able to further simplify the calculation by finding the coefficients in a model with only one scalar.  We carry out this calculation in the follow sections.

\section{A Single Fermion Toy model}
\label{sec:toymodel}
In our toy model, we have a single massless fermion and a single real scalar moving in a 4+1 dimensional spacetime. The Lagrangian is 
\begin{align}
\mathcal{L}&=\bar{\psi}(x,x_5)(i\gamma^\mu\partial_\mu-\gamma^5\partial_5-f\phi(x,x_5))\psi(x,x_5)+\frac{1}{2}\partial^\mu\phi(x,x_5)\partial_\mu\phi(x,x_5)\nonumber \\
&\quad-\frac{1}{2}\partial^5\phi(x,x_5)\partial_5\phi(x,x_5)-\frac{\lambda}{4}(\phi^2(x,x_5)-v^2)^2.
\label{eqn:lagrangian}
\end{align}

In order to obtain the chiral fermions, which we require to reproduce the SM, we compactify the extra dimension as an orbifold $S^1/\mathbf{Z}_2$, which imposes the boundary conditions: 
\begin{align}
\phi(x,-x_5)&=-\phi(x,x_5),&\phi(x,\pi R+x_5)&=-\phi(x,-\pi R +x_5),\nonumber \\
\psi(x,-x_5)&=\gamma_5\psi(x,x_5),& \psi(x,\pi R +x_5)&=\gamma_5\psi(x,-\pi R +x_5),
\label{eqn:OBCs}
\end{align}
 where $R$ is the radius of the extra dimension.

As with any theory where particles can propagate in compact extra dimensions, the effective 4d Lagrangian will contain a tower of KK modes:
\begin{equation}
\psi(x,x_5)=\sum^\infty_{n=0}\left(\psi_{n_R}(x)\xi_{n_R}(x_5)+\psi_{n_L}(x)\xi_{n_L}(x_5)\right),
\label{eqn:tower}
\end{equation}
where we have also performed a chiral decomposition.  This leads to a Dirac equation for the extra dimensional part of the wavefunction:
\begin{align}
(\partial_5+f\langle\phi(x_5)\rangle)\xi_{m_R}&=m\xi_{m_L},\nonumber\\
(-\partial_5+f\langle\phi(x_5)\rangle)\xi_{m_L}&=m\xi_{m_R}.
\label{eqn:DE}
\end{align}

Above the breaking scale, where $\langle\phi(x_5)\rangle=0$, these equations are easily solved and with the OBCs \eqref{eqn:OBCs} just give the usual KK chiral modes which we label $\chi_n$:
\begin{equation}
\spinor{\chi_{n_R}(x_5)}{\chi_{n_L}(x_5)}\equiv N_{\chi_n}\spinor{\cos nx_5/R}{\sin nx_5/R},\qquad m_{\chi_n}=\frac{n}{R}, \qquad n\in \mathbf{N}.
\label{eqn:unbrokensolns}
\end{equation}

At a temperature $T_\textrm{BAU}\sim 1/\pi R$ the scalar may undergo a phase transition and acquire a VEV.  It is at the boundary wall between these two phases that we envisage the BAU being generated. The OBCs, Eq. \eqref{eqn:OBCs}, requiring $\phi$ to be odd about the fixed points, lead to a double kink solution along the extra dimension, given approximately by \cite{hepph0007350}:
\begin{equation}
\langle\phi(x_5)\rangle=v \tanh\left(\sqrt{\frac{\lambda v^2}{2}}x_5\right)\tanh\left(\sqrt{\frac{\lambda v^2}{2}}(\pi R-x_5)\right)+\mathcal{O}(e^{-\pi R \sqrt{\lambda v^2}}).
\label{eqn:vev}
\end{equation}
Clearly this approximation is valid so long as the brane is much narrower the size of the extra dimension\footnote{An exact solution has been found for the scalar VEV profile (as well as its KK modes and those of the fermion) \cite{hepph0401108} and as expected it is in very good agreement with the approximation in this limit.} i.e.
\begin{equation}
\sqrt{\frac{1}{\lambda v^2}} \ll \pi R.
\label{eqn:large}
\end{equation}
We note that this is the same condition required for a split fermion model to effectively suppress Higgs Yukawa couplings and proton decay. In \cite{hepph0401070,hepph0505222} this scalar profile was approximated by a step function.

Combining equations \eqref{eqn:DE} and substituting in Eq. \eqref{eqn:vev} for the scalar profile we obtain two second order differential equations \cite{hepph0309115}:
\begin{align}
(-\partial_5^2+V_R)\xi_{m_R}&=m^2\xi_{m_R},\nonumber\\
(-\partial_5^2+V_L)\xi_{m_L}&=m^2\xi_{m_L},
\label{eqn:2ndorder}
\end{align}
where,
\begin{align}
V_{R,L}&=\mp uw\left(\frac{\tanh u(\pi R-x_5)}{\cosh^2ux_5}-\frac{\tanh ux_5}{\cosh^2u(\pi R-x_5)}\right)\nonumber\\
&\quad+w^2\tanh^2ux_5\tanh^2u(\pi R -x_5),
\label{eqn:Vdefs}
\end{align}
and,
\begin{equation}
w=fv, \quad u=\sqrt{\frac{\lambda v^2}{2}}.
\label{eqn:wu}
\end{equation}

In general, we can't solve these equations but in the narrow brane case the solutions can be related to the solutions about a scalar kink in an infinite extra dimension.  These solutions were derived in \cite{hepph0007350,hepph0309115} and we reproduce them in Appendix \ref{sec:infsolns}. In the following paragraphs we regularly refer to objects defined therein.

Note, in the narrow brane limit:
\begin{align}
V_R(x_5)\approx\left\{\begin{array}{ll}
V_R^{(\infty)}(x_5)\quad&\textrm{near }x_5=0,\\
V_L^{(\infty)}(x_5-\pi R)\quad&\textrm{near }x_5=\pi R,
\end{array}\right.\nonumber\\
V_L(x_5)\approx\left\{\begin{array}{ll}
V_L^{(\infty)}(x_5)\quad&\textrm{near }x_5=0,\\
V_R^{(\infty)}(x_5-\pi R)\quad&\textrm{near }x_5=\pi R,
\end{array}\right.
\label{eqn:approxVs}
\end{align}
thus we can approximate the solutions by those for the infinite case, Eqs. \eqref{eqn:boundsolns},\eqref{eqn:unboundsolns}.  As in the infinite case we find both bound solutions localized around the fixed points and unbound solutions which propagate throughout the extra dimension. However, the addition of OBCs further restricts the allowed solutions by requiring the right-handed (left-handed) solutions to be even (odd) about both fixed points.

\begin{figure}
\centering
\includegraphics[width=1\textwidth]{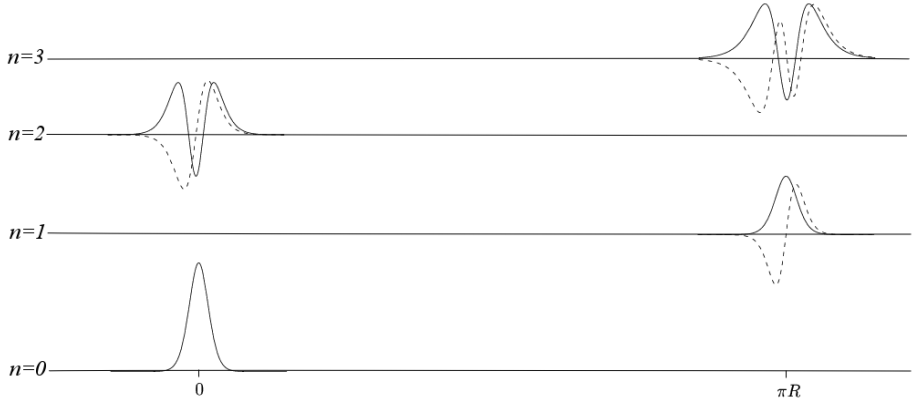}
  \caption{The bound modes, $m<w$, are localized about the fixed points of the extra dimension.  The even KK modes are found about $x_5=0$ while the odd modes are found at $x_5=\pi R$.  OBCs require that the right-handed modes (solid lines) be even functions about the fixed points, while the left-handed modes (dotted lines) must be odd.  There is no normalisable left-handed zero mode ensuring the SM derived from this model will be chiral as required.}
  \label{fig:tower}
\end{figure}

Thus, for the bound modes, when $n$ is even, $\xi_{n_R}(x_5)\approx\xi_{n_R}^{(\infty)}(x_5)$ around $x_5=0$ while the solution around $x_5=\pi R$ is suppressed since it would be odd i.e. $\xi_{n_R}(x_5)=0$ about $x_5=\pi R$.  Similarly, when $n$ is odd, the right-handed solution is suppressed around $x_5=0$ but takes the form $\xi_{n_R}(x_5)\approx\xi_{n_L}^{(\infty)}(x_5-\pi R)$ about $x_5=\pi R$.  The matching of the allowed solution to zero at the mid-point between the branes will be unproblematic since the solution will have already asymptoted to zero in the narrow brane context we are considering.  Similar arguments apply for the left-handed solution with there again being solutions localized to $x_5=0$ ($x_5=\pi R$) for $n$ even (odd) (although when $n=0$ there is no allowed solution, ensuring we get the chiral zero mode we require).  This is summarized graphically for the first few modes in Figure \ref{fig:tower}.  The expression for the masses of these modes carries straight over from the infinite case:
\begin{equation}
m^2_{\xi_n}=2nuw-n^2u^2,\quad \textrm{where }0\leq n<\frac{w}{u}.
\label{eqn:masses}
\end{equation}
Note that provided $w>u$ (i.e. $f>\sqrt{\frac{\lambda}{2}}$), there will be massive states localized at the fixed points.  Such modes were not present in the analysis of \cite{hepph0401070,hepph0505222} as their simplification of considering the scalar kink as a step function is equivalent to taking $\lambda\rightarrow \infty$ thereby removing these localized massive modes. Since the reflection asymmetries are essentially determined by the KK spectra in the two phases, this difference is potentially significant (especially since these modes are the lightest KK modes in the broken phase).

For the unbound modes, there exist both odd and even solutions of each chirality in the infinite extra dimension case (see Eq. \eqref{eqn:unboundsolns}) so we can always find a solution satisfying the OBCs about each fixed point:
\begin{align}
\zeta_{m_R}(x_5)&=\left\{\begin{array}{lr}
\zeta_{{m_R},\textrm{even}}^{(\infty)}(x_5),& 0\le x_5 \le \pi R/2,\\
\zeta_{{m_L},\textrm{even}}^{(\infty)}(x_5-\pi R),& \pi R/2 < x_5 \le \pi R, \end{array}\right.\\
\zeta_{m_L}(x_5)&=\left\{\begin{array}{lr}
\zeta_{{m_L},\textrm{odd}}^{(\infty)}(x_5),&  0\le x_5 \le \pi R/2,\\
\zeta_{{m_R},\textrm{odd}}^{(\infty)}(x_5-\pi R),& \pi R/2 < x_5 \le \pi R. \end{array}\right.
\end{align}
Continuity of the wavefunction means we must match these solutions at the mid-point between the fixed points, $x_5=\pi R/2$.  For the right- and left-handed solutions respectively, this gives:
\begin{align}
\zeta_{{m_R},\textrm{even}}^{(\infty)}(\pi R/2)&=\zeta_{{m_L},\textrm{even}}^{(\infty)}(-\pi R/2),\\
\zeta_{{m_L},\textrm{odd}}^{(\infty)}(\pi R/2)&=\zeta_{{m_R},\textrm{odd}}^{(\infty)}(-\pi R/2).\label{eqn:match}
\end{align}
In order that such a matching can be done in general, the phases of the two wavefunctions must match, which can always be achieved by an appropriate choice of normalization constants.  The matching only occurs for certain values of $m_{\zeta_n}$, leading to a discretized tower of allowed unbound modes. Note these matching conditions differ from those found in \cite{hepph0309115}.  In general we cannot solve these equations analytically and we must resort to numerical techniques.  Having done so, we will have an infinite tower of modes of each chirality which we call $\zeta_{n_{R,L}}$.

\section{Reflection Coefficients and Baryon Asymmetry}
\label{sec:coeffs}
Having found the solutions to the Dirac equations in both the broken and unbroken phases, we now proceed to calculate the reflection amplitudes for the scattering of incoming KK fermions off the phase boundary.  This involves matching the wavefunctions of the fermions in the broken and unbroken phases on the boundary between them, in exact analogy with the procedure for EW baryogenesis described in Section \ref{sec:EWB}. As in that case, we may Lorentz transform away any components of the momentum of the incoming particle along uncompactified dimensions parallel to the wall (that is, along the $x_1$ and $x_2$ directions if we define the $x_3$-axis to be normal to the wall) and position the wall at $x_3=0$.  We cannot, however, transform away momenta along the extra dimension since here Lorentz invariance is broken by the orbifold fixed points.  As a result we are matching the wavefunctions not only at a point on the boundary but right around the extra dimension.

We consider in turn an incoming particle of each KK mode, $\chi_{n'}$, with helicity $h=\pm 1$.  This can be reflected back in to the unbroken phase as $\chi_n$ with amplitude $r_n$, or transmitted to the broken phase as $\zeta_n$ or $\xi_n$ with amplitudes $t_{\zeta_n}$ and $t_{\xi_n}$ respectively.  The wavefunction matching condition on the bubble wall is then:
\begin{align}
\left\{\sum_{n=0}^\infty\left[ r_ne^{-ik_3^{\chi_n}x_3}\spinor{\sqrt{E-hk^{\chi_n}_3}\chi_{n_R}(x_5)}{\sqrt{E+hk^{\chi_n}_3}\chi_{n_L}(x_5)}\right]\right.\qquad\qquad\qquad\qquad\qquad&\nonumber\\
+\left.\left.e^{-ik_3^{\chi_{n'}}x_3}\spinor{\sqrt{E+hk_3^{\chi_{n'}}}\chi_{n'_R}(x_5)}{\sqrt{E-hk_3^{\chi_{n'}}}\chi_{n'_L}(x_5)}\right\}e^{-iEt}\right|_{x_3=0}&\nonumber\\
=\left\{\sum_{n=1}^\infty \left[t_{\zeta_n}e^{-ik_3^{\zeta_n}x_3} \spinor{\sqrt{E+hk_3^{\zeta_n}}\zeta_{n_R}(x_5)}{\sqrt{E-hk_3^{\zeta_n}}\zeta_{n_L}(x_5)}\right]\right.\qquad\qquad\qquad\qquad\qquad&\nonumber\\
+\left.\left.\sum_{n=0}^{\lfloor w/u\rfloor}\left[t_{\xi_n} e^{-ik_3^{\xi_n}x_3}\spinor{\sqrt{E+hk_3^{\xi_n}}\xi_{n_R}(x_5)}{\sqrt{E-hk_3^{\xi_n}}\xi_{n_L}(x_5)}\right]\right\}e^{-iEt}\right|_{x_3=0},&
\label{eqn:spectrum}
\end{align}
where $\lfloor y\rfloor$ is the largest integer smaller than $y$, $E$ is the energy of the incoming particle and $k_3^{\chi_n,\zeta_n,\xi_n}=\sqrt{E^2-(m_{\chi_n,\zeta_n,\xi_n})^2}$ are the particle momenta in the $x_3$-direction where we recall that the masses must be found numerically for the $\zeta_n$s.

To solve this equation, we find Fourier series expansions for the $\xi_n$ and $\zeta_n$. We may then equate the Fourier coefficients on the two sides term by term.  Since this must be done numerically, it is necessary to truncate the infinite sums in Eq. \eqref{eqn:spectrum} at $N^\chi$ and $N^\zeta$ respectively.  For sufficiently high truncation, this leaves us with an over-determined set of simultaneous equations in the $r$s and $t$s which we may solve by the pseudo-inverse method.

\begin{figure}[htbp]
  \centering
  \includegraphics[width=0.8\textwidth]{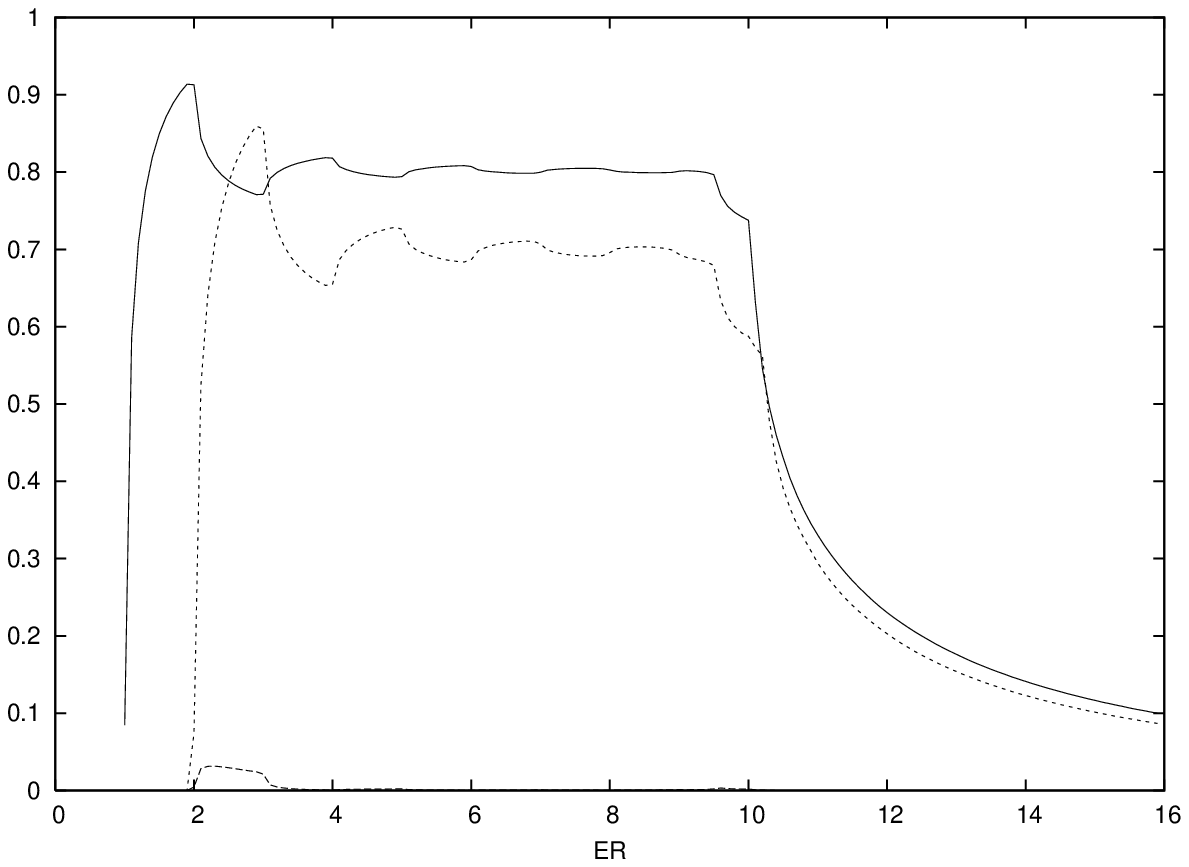}\\
  \includegraphics[width=0.8\textwidth]{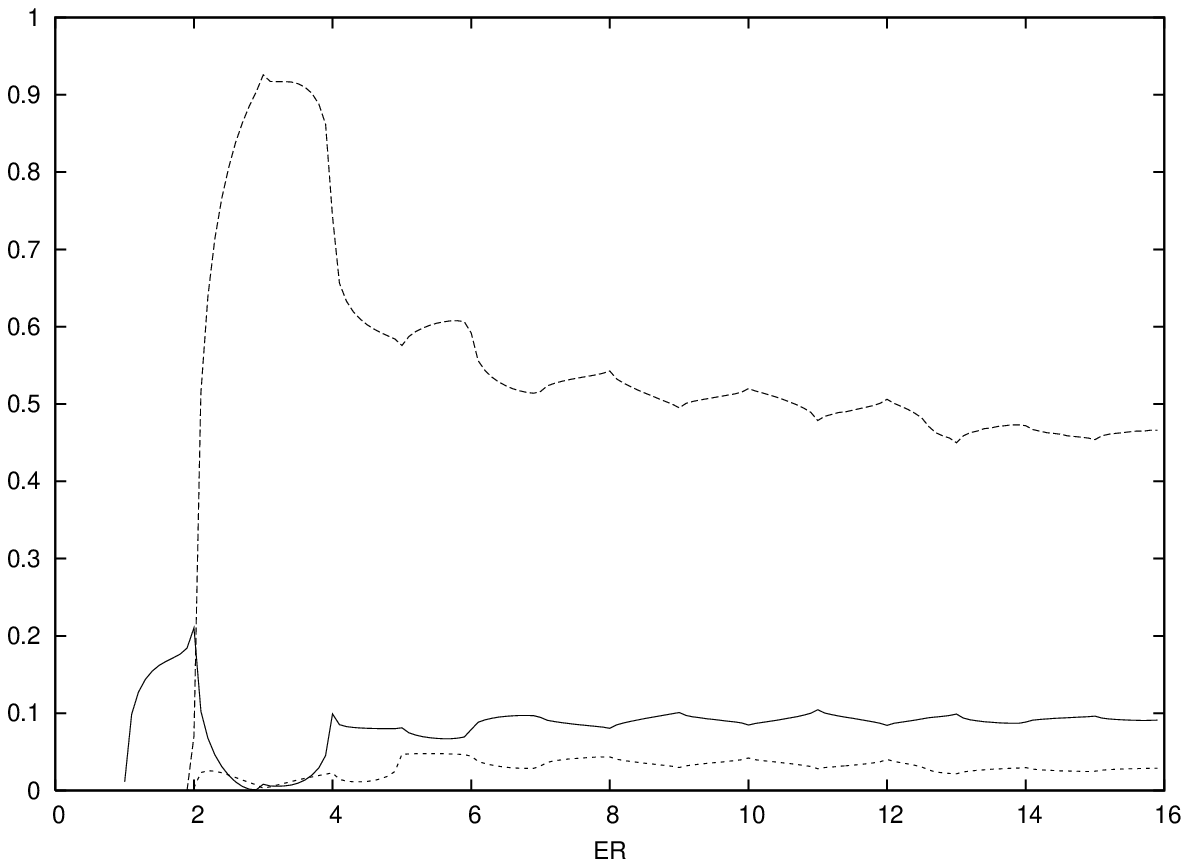}
  \caption{Plots of $\mathcal{R}^{1,0}$ (solid line), $\mathcal{R}^{2,0}$ (dashed line) and $\mathcal{R}^{2,1}$ (dotted line) as a function of the incoming particle's energy for $v=10/R, f=1, \lambda=1$ (top plot) and $v=20/R, f=1, \lambda=.1$ (bottom plot).}
  \label{fig:Rnm}
\end{figure}

We can then find the reflection and transmission coefficients:
\begin{align}
\mathcal{R}^{nn'}(E)&=\textrm{Re}\left(\frac{k^{\chi_n}_3}{k^{\chi_{n'}}_3}\right)|r_n|^2,\\
\mathcal{T}^{nn'}_\zeta(E)&=\textrm{Re}\left(\frac{k^{\zeta_n}_3}{k^{\chi_{n'}}_3}\right)|t_{\zeta_n}|^2,\\
\mathcal{T}^{nn'}_\xi(E)&=\textrm{Re}\left(\frac{k^{\xi_n}_3}{k^{\chi_{n'}}_3}\right)|t_{\xi_n}|^2.
\label{eqn:RTcoeffs}
\end{align}
Unitarity should ensure that
\begin{equation}
\mathcal{R}^{n'}(E)+\mathcal{T}^{n'}(E)\equiv \sum_{n=0}^{N^\chi}\mathcal{R}^{nn'}+\left(\sum_{n=1}^{N^\zeta}\mathcal{T}^{nn'}_\zeta+\sum_{n=0}^{\lfloor w/u \rfloor}\mathcal{T}^{nn'}_\xi\right)=1,
\end{equation}
for all $n'$ and $E$ and this is indeed the case provided we choose $N^{\chi,\zeta}$ sufficiently large.  Plots of $\mathcal{R}^{nn'}$ as a function of the incoming particle's energy are shown in Figure \ref{fig:Rnm} for two choices of values for the parameters $f, \lambda, v$. In these plots the effect on the reflection coefficients of additional modes coming on-shell as the incoming particle's energy is increased is clearly visible.  It is also apparent that the choice of parameters has a significant impact on the values the various reflection coefficients take, suggesting the value we find for $n_B/s$ may be strongly parameter dependent.

However, when we use Eq. \eqref{eqn:npert} to calculate the BAU we find $n_B/s\sim 10^{-5}\epsilon_{B-L}$, relatively independently of the choice of values for the parameters, provided we remain in the narrow brane regime of Eq. \eqref{eqn:large}.  Further, as long as $T_{\textrm{BAU}}$ is of the same order as $1/\pi R$, its precise value does not change the order of magnitude of the resulting asymmetry. This number is a couple orders of magnitude smaller than that found in similar calculations \cite{hepph0401070,hepph0505222} where the scalar kink is taken as a step function and there are no localized massive modes in the broken phase\footnote{Although \cite{hepph0505222} assumes the necessary CP violation comes from the CKM phase of the SM, while we assume it is from the extended scalar sector, a comparison of results is still possible since in both cases the violation is $\mathcal{O}(1)$.}, confirming that these extra modes have a significant effect on the result.  However, it is of the same order of magnitude as another calculation of $n_B/s$ in \cite{hepph0505222}. This calculation considers a three generation model, although it does not implement a quark-lepton separation to suppress proton decay. It uses a step function as the scalar profile and uses a different method to calculate the reflection asymmetries, exploring a different region of the parameter space, namely where $v\ll 1/R$. Interestingly, a small $v$ would require us to take $\lambda$ large to remain in the narrow brane regime.  Recall that taking $\lambda$ to be large is precisely the limit in which our choice of scalar profile tends towards a step function meaning the choice of scalar profile may not be as crucial in this region of parameter space.

In order that this mechanism generate the known value of $n_B/s\sim 10^{-10}$ requires $\epsilon_{B-L}\sim10^{-5}$ which by Eq. \eqref{eqn:epsilonB-L} gives
\begin{equation}
M_*\sim \frac{10}{\pi R}.
\end{equation}
Given the relative simplicity of our construction and the NDA used to derive the expression for $\epsilon_{B-L}$, one should perhaps not take this relation too seriously.  However, it does suggest that in order for split fermion baryogenesis to be viable the fundamental energy scale, $M_*$, at which $B-L$ violating theory applies can not be orders of magnitude higher than the scale, $T_{\textrm{BAU}}$, at which fermion localization occurs and the BAU is generated.

The major uncertainty in our calculation is the degree to which our single fermion result carries over to a realistic three generation model.  The issues here can be separated into two types.  Firstly, there are those fundamental to going to a three generation model -- to what extent does mixing between different fermion flavors change the reflection coefficients?  On this front it is suggestive that \cite{hepph0505222} find a value for $n_B/s$ in a three generation calculation which agrees with our result, albeit in a very different region of parameter space.  The second issue is not fundamental to a three generation model but results from what else we are requiring the spliting of the fermions to provide -- a suppression of proton decay and an origin for the hierarchy in Higgs Yukawa couplings.  As discussed early, this requires the various fermions to be localized to different points in the extra dimensional bulk, with quarks separated from leptons (to suppress proton decay) and $SU(2)_L$ doublets from singlets (to provide a Yukawa hierarchy).  Having fermions not only localized at the fixed points but also in the bulk may have some important effects.  Unfortunately, there has yet to be a realistic split fermion model which dynamical localizes the fermions in such a way as to achieve both these outcomes.  It may be that some additional physics is required to achieve this and until we have such a dynamical realization of the idea, we are unable to cosider how well our toy model result generalizes to it.

\section{Conclusion}
\label{sec:conclusion}
Split fermion models are attractive extra dimensional constructions which have the potential to explain both the hierarchy in SM fermion masses and the suppression of proton decay.  In this paper, we have carried out a calculation to estimate the baryon asymmetry generated in such models during the phase transition where the localizing scalar acquires a VEV. We have improved on previous similar calculations by considering a more realistic localizing scalar profile and resultant KK spectrum.  For a range of parameters and transition temperatures, we find $n_B/s\sim 10^{-5}\epsilon_{B-L}$ compared with a known value of $n_B/s\sim 10^{-10}$. This value is a couple orders of magnitude smaller than the past calculations but still consistent with this mechanism being the source of the BAU provided the scale of $B-L$ violating physics is not too much higher than that of the fermion localizing physics.  We have made numerous simplifications the most significant of which is calculating the reflection coefficients in a single fermion framework and anticipating that the reflection coefficients found will carry over to three generation models.  It is interesting that our result for the generated BAU is of the same order as a three generation perturbative calculation \cite{hepph0505222} but this calculation is done in a very different region of parameter space and does not separate quarks and leptons as a realistic model must. While our results are suggestive, clearly more work is necessary -- both to construct more realistic dynamical implementations of the original split fermion model and in refining our calculations to better include all the relevant physics -- in order to draw any firm conclusions as to whether this mechanism provides a viable explanation of the matter-antimatter asymmetry of the universe.

\subsection*{Acknowledgements}
I would like to thank Girish Joshi and Damien George for useful discussions and their helpful comments.

\appendix
\section{Split Fermion Kaluza-Klein modes in an infinite extra dimension}
\label{sec:infsolns}

The Kaluza-Klein modes for a fermion in the presence of a scalar kink in an infinite extra dimension have previously been calculated by \cite{hepph0007350,hepph0309115}.  Here we quote the key results.

In the case of an infinite extra dimension the scalar VEV is,
\begin{equation}
\langle\phi(x_5)\rangle=v \tanh\left(\sqrt{\frac{\lambda v^2}{2}}x_5\right),
\label{eqn:infvev}
\end{equation}
where we now have only a single fixed point at $x_5=0$.  Proceeding as we did in Section \ref{sec:toymodel}, we derive second order differential equations,
\begin{align}
(-\partial_5^2+V_R^{(\infty)})\xi_{m_R}&=m^2\xi_{m_R},\nonumber\\
(-\partial_5^2+V_L^{(\infty)})\xi_{m_L}&=m^2\xi_{m_L},
\label{eqn:inf2ndorder}
\end{align}
where,
\begin{equation}
V_{R,L}^{(\infty)}=\mp uw\frac{1}{\cosh^2ux_5}+w^2\tanh^2ux_5.
\label{eqn:infV}
\end{equation}

With some changes of variables, these equations can be written in the form of the standard hypergeometric differential equation \cite{HigherTransFns} which can then be solved.  There exist both bound ($m^2<w^2$) and unbound ($m^2>w^2$) solutions.  The bound solutions are found to be:
\begin{align}
\xi_{m_R}^{(\infty)}(x_5)&=N_{m_R}\left(\frac{1}{\cosh ux_5}\right)^\epsilon {}_2F_1\left(\epsilon-\frac{w}{u},\epsilon+\frac{w}{u}+1,\epsilon+1;\frac{1}{1+e^{2ux_5}}\right),\nonumber\\
\xi_{m_L}^{(\infty)}(x_5)&=N_{m_L}\left(\frac{1}{\cosh ux_5}\right)^\epsilon {}_2F_1\left(\epsilon-\frac{w}{u}+1,\epsilon+\frac{w}{u},\epsilon+1;\frac{1}{1+e^{2ux_5}}\right),
\label{eqn:boundsolns}
\end{align}
where $N_{m_{R,L}}$ are normalization constants, ${}_2F_1(a,b,c;z)$ is the hypergeometric function \cite{HigherTransFns} and we have defined
\begin{equation}
\epsilon=\sqrt{\frac{w^2-m^2}{u^2}}.
\end{equation}
In order that these solutions be normalizable, we have the additional condition that
\begin{equation}
m^2_n=2nuw-n^2u^2,\quad \textrm{where }0\leq n<\frac{w}{u},
\label{eqn:infmasses}
\end{equation}
leading to a discretized tower of allowed modes.  For even $n$, $\xi_R$ is even while $\xi_L$ is odd. For odd $n$, the reverse is true.  Further, when $n=0$, only the right-handed mode is normalizable meaning we obtain the chiral massless fermion we require if we hope to reproduce the SM.

For the unbound case, there are two linearly independent solutions for each chirality, which we may conveniently write as even and odd wavefunctions,
\begin{align}
\zeta_{m_R,\textrm{even}}^{(\infty)}(x_5)&=N'_{{m_R},\textrm{even}}\left[\left(\frac{1}{\cosh ux_5}\right)^\epsilon {}_2F_1\left(\epsilon-\frac{w}{u},\epsilon+\frac{w}{u}+1,\epsilon+1;\frac{1}{1+e^{2ux_5}}\right)\right.\nonumber \\
&\quad\left.+\frac{D_1}{1-D_2}(2e^{ux_5})^\epsilon {}_2F_1\left(-\frac{w}{u},\frac{w}{u}+1,-\epsilon+1;\frac{1}{1+e^{2ux_5}}\right)\right],\nonumber\\
\zeta_{m_R,\textrm{odd}}^{(\infty)}(x_5)&=N'_{{m_R},\textrm{odd}}\left[\left(\frac{1}{\cosh ux_5}\right)^\epsilon {}_2F_1\left(\epsilon-\frac{w}{u},\epsilon+\frac{w}{u}+1,\epsilon+1;\frac{1}{1+e^{2ux_5}}\right)\right.\nonumber \\
&\quad\left.-\frac{D_1}{1+D_2}(2e^{ux_5})^\epsilon {}_2F_1\left(-\frac{w}{u},\frac{w}{u}+1,-\epsilon+1;\frac{1}{1+e^{2ux_5}}\right)\right],\nonumber\\
\zeta_{m_L,\textrm{even}}^{(\infty)}(x_5)&=N'_{{m_L},\textrm{even}}\left[\left(\frac{1}{\cosh ux_5}\right)^\epsilon {}_2F_1\left(\epsilon-\frac{w}{u}+1,\epsilon+\frac{w}{u},\epsilon+1;\frac{1}{1+e^{2ux_5}}\right)\right.\nonumber \\
&\quad\left.+\frac{D_3}{1+D_2}(2e^{ux_5})^\epsilon{}_2F_1\left(\-\frac{w}{u}+1,\frac{w}{u},-\epsilon+1;\frac{1}{1+e^{2ux_5}}\right)\right],\nonumber\\
\zeta_{m_L,\textrm{odd}}^{(\infty)}(x_5)&=N'_{{m_L},\textrm{odd}}\left[\left(\frac{1}{\cosh ux_5}\right)^\epsilon {}_2F_1\left(\epsilon-\frac{w}{u}+1,\epsilon+\frac{w}{u},\epsilon+1;\frac{1}{1+e^{2ux_5}}\right)\right.\nonumber\\
&\quad\left.-\frac{D_3}{1-D_2}(2e^{ux_5})^\epsilon{}_2F_1\left(\-\frac{w}{u}+1,\frac{w}{u},-\epsilon+1;\frac{1}{1+e^{2ux_5}}\right)\right],
\label{eqn:unboundsolns}
\end{align}
where,
\begin{equation*}
D_1=\frac{\Gamma(\epsilon)\Gamma(\epsilon+1)}{\Gamma(\epsilon-w/u)\Gamma(\epsilon+w/u+1)}, \quad D_2=\frac{\Gamma(\epsilon)\Gamma(1-\epsilon)}{\Gamma(w/u+1)\Gamma(-w/u)},
\end{equation*}
\begin{equation}
D_3=\frac{\Gamma(\epsilon)\Gamma(\epsilon+1)}{\Gamma(\epsilon-w/u+1)\Gamma(\epsilon+w/u)}.
\label{eqn:Ds}
\end{equation}
In this case, there is no restriction on $m$ and there is a continuum of modes.  As we would expect, as $x_5\to \infty$, these tend towards the usual plane wave solutions.

\bibliographystyle{h-physrev3}
\bibliography{bibliography}
\end{document}